\documentclass[10pt,twocolumn,twoside,letterpaper]{IEEEtran}

\usepackage{geometry}
\makeatletter

\usepackage{url}

\usepackage{cite}
\usepackage{amsmath,amssymb,amsfonts}
\usepackage{graphicx}
\usepackage{textcomp,nicefrac}
\usepackage{subcaption}
\usepackage{hhline}
\usepackage[dvipsnames]{xcolor}
\usepackage[final]{changes}
\begin{document}
\title{Enabling PSD-capability for a High-density Channel Imager}
%
%
%

\author{Ming Fang, 
	   Satwik Pani, 
	   and Angela Di Fulvio
\thanks{Manuscript received December 1, 2021. This work is funded in-part by the Nuclear Regulatory Commission Faculty Development Grant number 31310019M0011.}
\thanks{M. Fang, S. Pani, and A. Di Fulvio are with the Department of Nuclear, Plasma, and Radiological Engineering, University of Illinois at Urbana-Champaign, Urbana, IL 61801, United States (telephone: 217-305-1769, e-mail: mingf2@illinois.edu;pani2@illinois.edu;difulvio@illinois.edu).}}

\maketitle

\pagenumbering{gobble}

\begin{abstract}
Pulse shape discrimination (PSD) is crucial for non-proliferation and security applications, where fast neutrons need to be identified and measured in the presence of a strong gamma ray background. The traditional charge-integration based PSD method requires the storage and processing of hundreds of  samples  for  each  single  pulse, which is time- and memory-consuming for high density channel applications. In this work, we explored the possibility of implementing PSD using a commercial ASIC that allows the user to adjust the pulse shaping time. We demonstrated that PSD can be achieved by by maximizing the difference between the pulse shaping circuit's responses to neutron and gamma ray pulses. 
\end{abstract}


\section{Introduction}
%
%
%
%
\IEEEPARstart{M}{any} scintillating materials exhibit PSD capability that enables to discriminate different types of radiation-produced light pulses, such as gamma-ray and neutron pulses, based on their time-dependent shape. It is crucial to implement PSD in imaging applications, where fast neutrons need to be detected and identified, often in the presence of a strong gamma ray background. Traditional approaches rely on the digitization of detected scintillation pulses and extraction of a pulse shape discrimination parameter that enables this functionality. However, this approach requires the storage and processing of hundreds of samples for each pulse. This process is typically performed off-line and its implementation in real time is prohibitive for time consumption and memory requirement. Furthermore, it is critical when dealing with tens of channels as in high density scatter cameras. An alternative approach relies on using Application Specific Integrated Circuits (ASIC) for the fast processing of silicon photomultiplier (SiPM) pulses. Commercial ASICs do not typically implement PSD functionality. In this work, we have designed and built a high-density channel PSD-capable imager, based on CsI(Tl) and PSD capable EJ-276 plastic scintillator. The light readout is performed by SiPM matrices on both sides of the detectors. We used a commercial ASIC (Citiroc1A by Weeroc) and demonsrated via simulation the possibility of implementing PSD by using different pulse shaping time constants of the front-end input stage.
\section{Methods}
\subsection{Imager System Design}
\begin{figure}[!htbp]
    \centering
    \includegraphics[width=\linewidth]{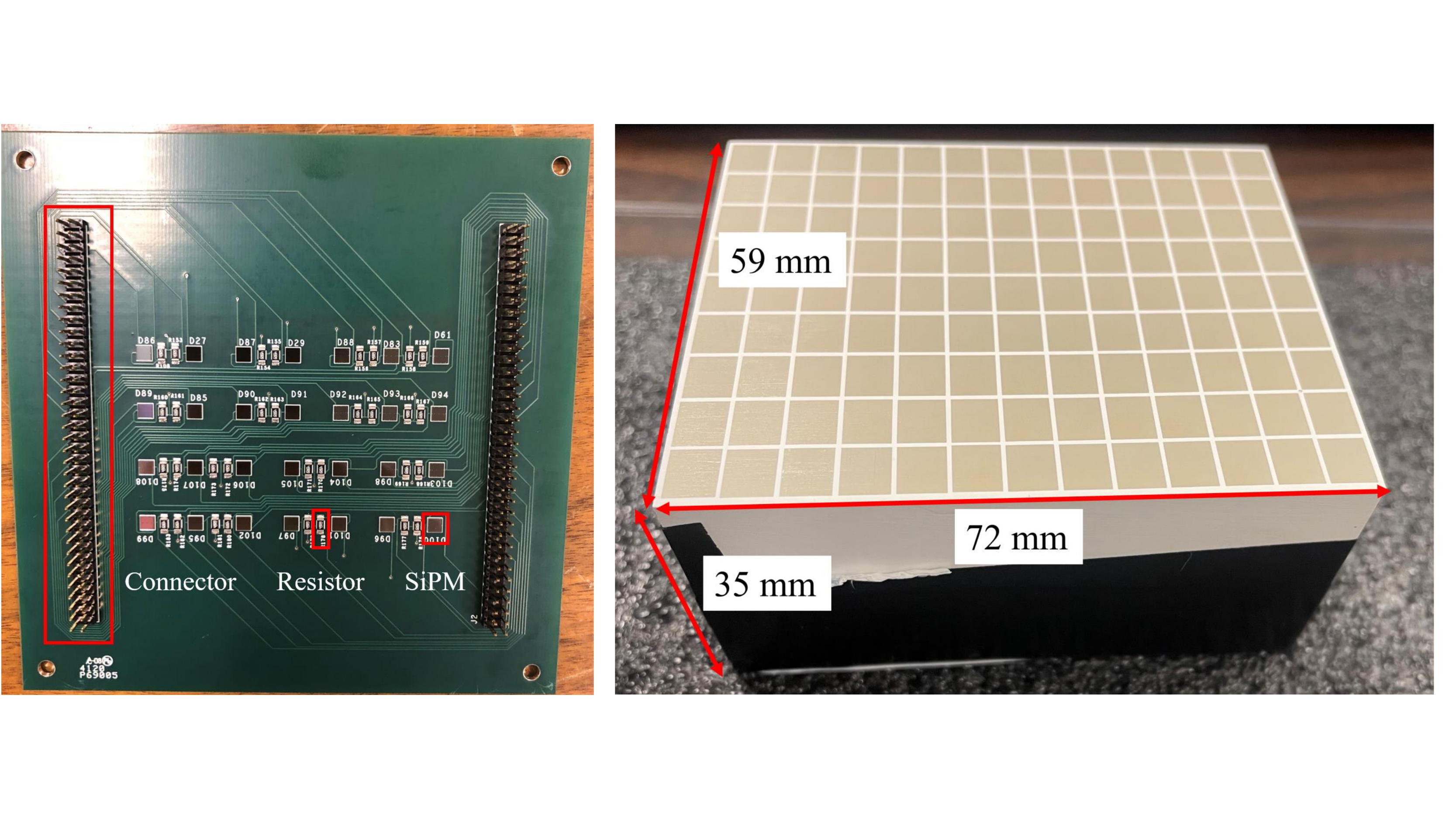}
    \caption{Left: custom PCB hosting a SiPM array. Right: a pixelated CsI(Tl) crystal.}
    \label{fig:imager2}
\end{figure}

We designed and built two 12~cm × 12 cm~custom PCB, each hosting a 4 × 7 SenSL MicroFJ-30020 SiPM~\cite{semiconductor2017silicon} array, which are coupled to the top and bottom surfaces of a pixelated CsI(Tl) crystal to collect the scintillation light pulses (Fig.~\ref{fig:imager2}). The SiPM array covers an area of 7~cm × 4.3~cm on the scintillator surface. The vertical and horizontal distance between two neighboring SiPMs is 13~mm and 11~mm, respectively. 

\begin{figure}[!htbp]
    \centering
    \includegraphics[width=\linewidth]{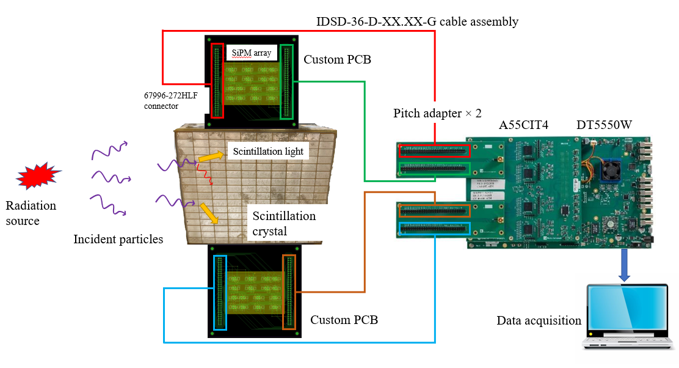}
    \caption{Schematic design of the imaging system.}\label{fig:imager1}
     \vspace{-15pt}
\end{figure}

Fig.\ref{fig:imager1} shows the schematic design of the imaging system. There are two double-row 2.54 mm-pitch 72 pin connectors on the custom PCBs, which provide the biasing voltages and transfer the output signals of the SiPMs. The connector is connected to the A55CIT4 + DT5550W readout system by CAEN~\cite{a55citx2020}\cite{dt5550w2020} through a pitch adapter kit provided with the readout system. The A55CIT4 board hosts four Citiroc1A  Weeroc ASICs, each integrating 32 readout channels. Thus, there are 128 bias/signal couples available in total. The CAEN A7585D Power Supply module on A55CIT4 provides the bias voltages for the SiPMs. The readout system is controlled by the SCI-5550W Readout Software on a lab computer, which allows us to adjust the acquisition settings including pulse shaping times of each channel.


\subsection{PSD}\label{section:PSD}

\begin{figure}[!htbp]
    \centering
    \includegraphics[width=\linewidth]{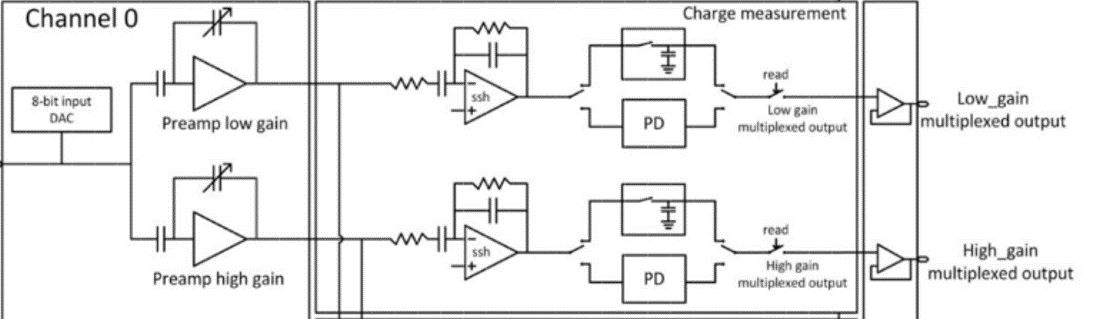}
    \caption{Schematic of the readout electronics on Citiroc1A ASIC~\cite{a55citx2020}.}
    \label{fig:schematic}
\end{figure}

Charge-integration based PSD is both resource-intensive and slow for real-time applications such as imaging, when large number of channels are involved. We investigated the possibility of using available electronics on Citiroc1A ASIC to perform PSD. The readout electronics of a single channel are shown in Fig.~\ref{fig:schematic}. The input signal is split into two paths, referred to as the low gain (LG) path and high gain (HG) path. The LG and HG signals then pass through a filter with independently adjustable pulse shaping times. Two independent output signals are produced and the amplitudes of the output signals are reported, referred to as the LG output and HG output. Since neutron and gamma ray pulses have different pulse shapes and respond differently to the filter, it is possible to perform PSD by tuning the shaping times and maximizing the difference between filtered pulse's height from neutron and gamma ray pulses. 

\begin{figure}[!htbp]
    \centering
    \includegraphics[width=\linewidth]{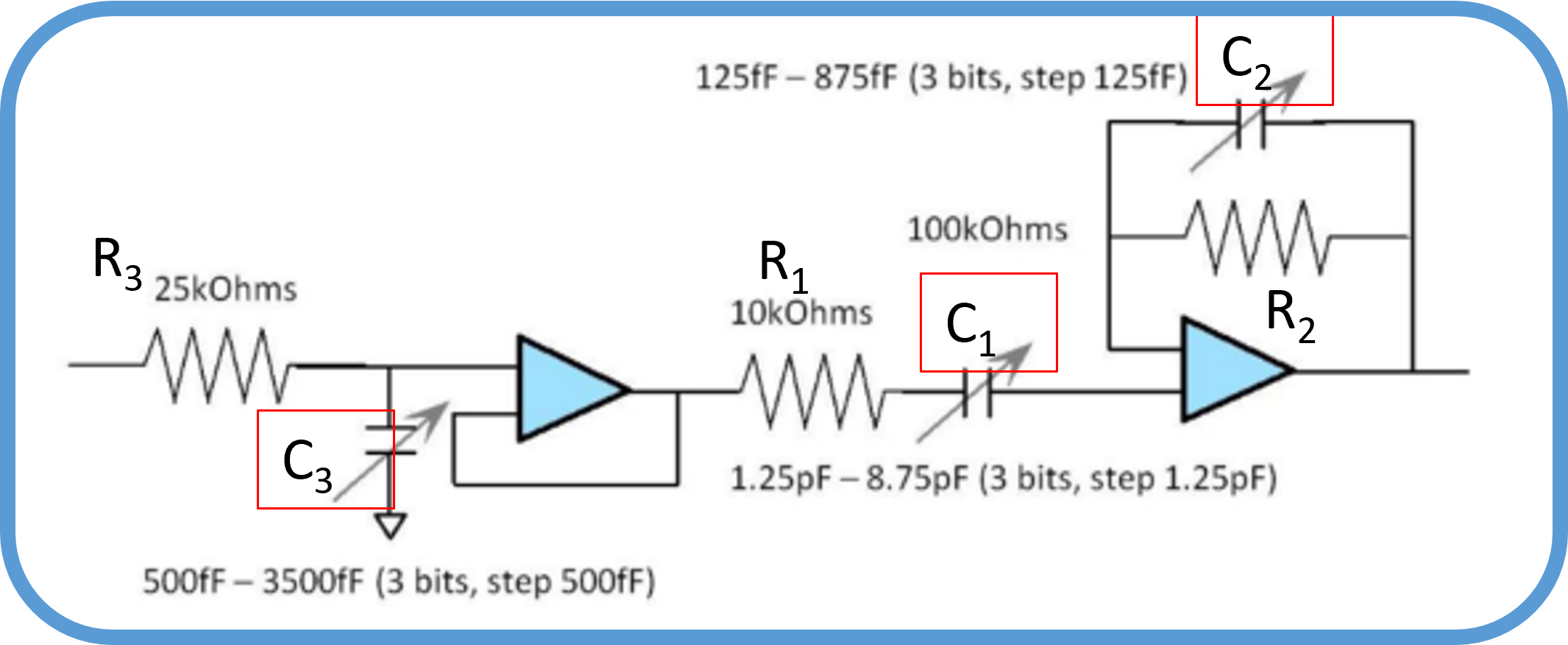}
    \caption{Schematic of the pulse shaper on Citiroc1A ASIC~\cite{a55citx2020}. $R_1 = 10~k\Omega, R_2 = 100~k\Omega, R_3 = 25~k\Omega$. $C_1, C_2, C_3$ are controlled by an user-input parameter, \textit{peaking time}, as shown in Table~\ref{table:peaking_times}. }
    \label{fig:shaper}
\end{figure}

Fig.~\ref{fig:shaper} shows the pulse filter that LG/HG signal passes through on the Citiroc1A ASIC. The A55CIT4 board allows the user to vary $C_1 , C_2 , C_3$ together through a “peaking time” parameter, thus changing the frequency response of the filter. The frequency-dependent transfer function (FDTF) of the filter is:
\begin{equation}
\tiny
T(s)=\left(\frac{1}{s R_{3} C_{3}+1}\right)\left(\frac{s^{2} R_{1} C_{1} R_{2} C_{2}+s\left(R_{1} C_{1}+R_{2} C_{2}+R_{2} C_{1}\right)+1}{s^{2} R_{1} C_{1} R_{2} C_{2}+s\left(R_{1} C_{1}+R_{2} C_{2}\right)+1}\right)\label{eq:fdtf}
\end{equation}
\begin{equation*}
    s = i\omega, \omega = 2\pi f
\end{equation*}
where $T$ is the ratio between the input signal amplitude and the out signal amplitude, $f$ is the frequency.\added{ We simulated the FDTF in Eq.~\eqref{eq:fdtf} using python's Scipy package, which allows to calculate the LG/HG output at a given peaking time configuration for any input signals.}

\begin{table}[!ht]
    \caption{Configurable $C_1, C_2, C_3$ Values.}~\label{table:peaking_times}
    \centering
    \begin{tabular}{|c|c|c|c|}
    \hline
        \bfseries $\mathbf{C_1}$ (pF) & \bfseries $\mathbf{C_2}$ (fF) & \bfseries $\mathbf{C_3}$ (fF) & \bfseries Peaking time (ns) \\ \hhline{|=|=|=|=|}
        1.25 & 125 & 500 & 87.5 \\ \hline
        2.5 & 250 & 1000 & 75 \\ \hline
        3.75 & 375 & 1500 & 62.5 \\ \hline
        5 & 500 & 2000 & 50 \\ \hline
        6.25 & 625 & 2500 & 37.5 \\ \hline
        7.5 & 750 & 3000 & 25 \\ \hline
        8.75 & 875 & 3500 & 12.5 \\ \hline
    \end{tabular}
\end{table}

\begin{figure}[!htbp]
    \centering
    \includegraphics[width=\linewidth]{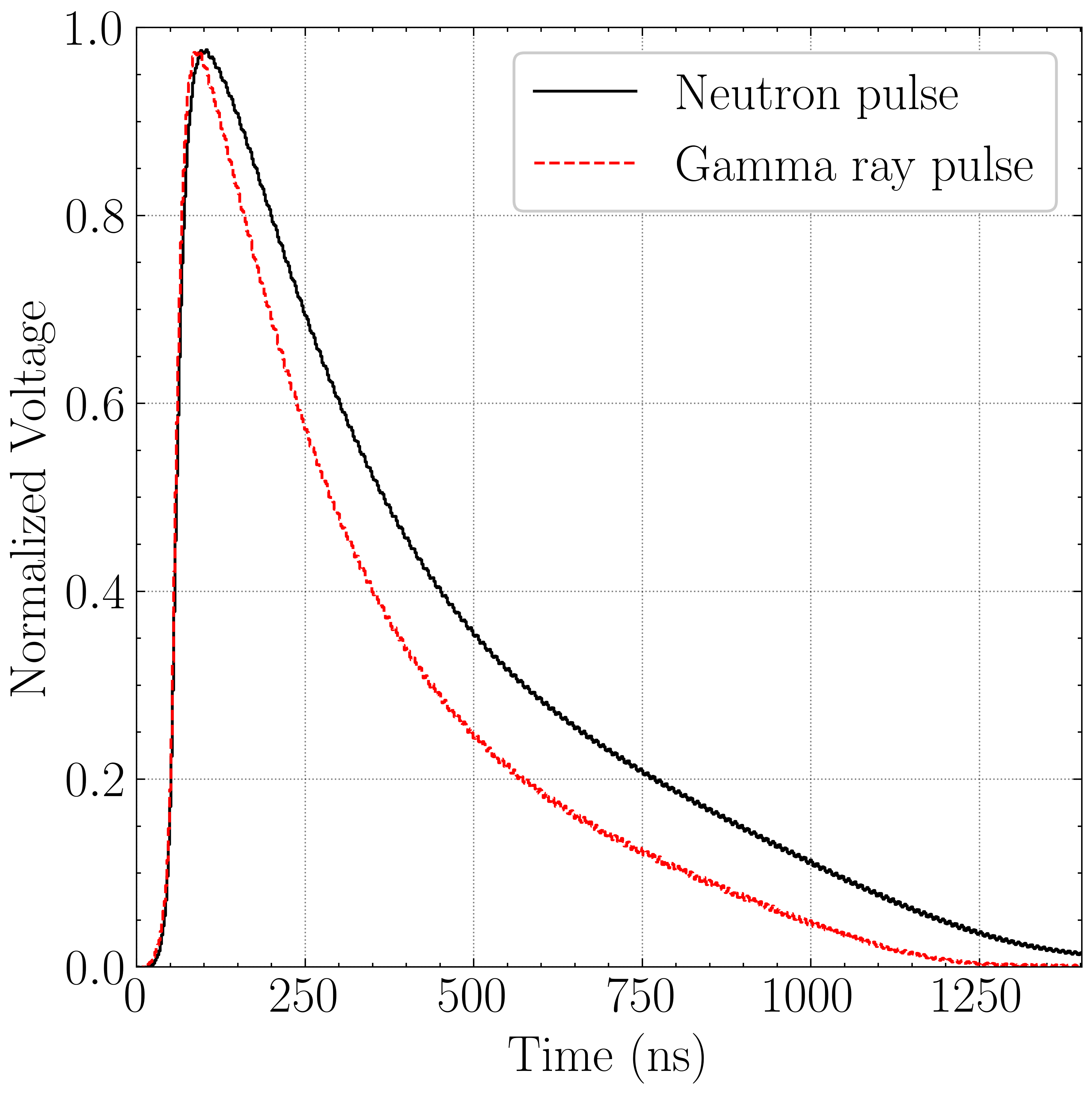}
    \caption{Normalized neutron and gamma ray pulses.}
    \label{fig:pulses}
\end{figure}

The PSD parameter is defined as the ratio between the LG output and HG output. To demonstrate the feasibility of our approach, we input a measured dataset consisting of 90,000 neutron/gamma ray pulses to the filter and calculated the PSD parameter for each pulse. These pulses were acquired by a 2’’x2’’ regular stilbene detector coupled with an array of four MicroFC-60035 SiPMs measuring a Cf-252 source. Pulses were digitized using a 14-bit 500MS/s DT5730 digitizer. Fig.~\ref{fig:pulses} shows a normalized neutron and gamma-ray pulse in this dataset. We optimized the LG and HG shaping times to obtain the best separation between the neutron and gamma ray pulses.

\section{Results}
\subsection{Gamma ray response}
\begin{figure}[!htbp]
    \centering
    \includegraphics[width=\linewidth]{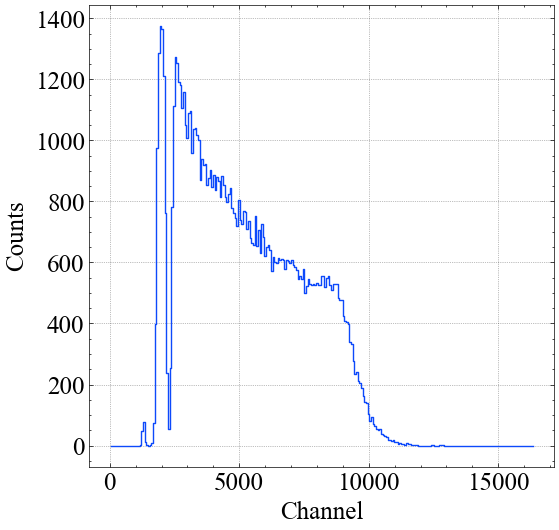}
    \caption{Cs-137 spectrum measured with an EJ-276 organic scintillator and A55CIT4+DT5550W readout system.}
    \label{fig:cs137_raw}
\end{figure}

\begin{figure}[!htbp]
    \centering
    \includegraphics[width=\linewidth]{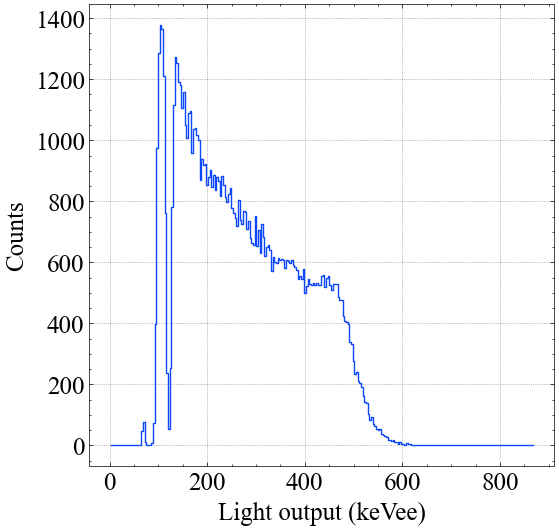}
    \caption{Calibrated Cs-137 light output spectrum.}
    \label{fig:cs137}
\end{figure}

We coupled a $6~\mathrm{ mm}\times 3~\mathrm{ mm}\times 3~\mathrm{ mm}$ EJ-276 organic scintillator to a MicroFJ-30020 SiPM. We measured a 30~$\mu$Ci Cs-137 source for 15 minutes using the system. The SiPM bias voltage was set to 28.5V and the LG peaking time and HG shaping time were set to 12.5 ns and 87.5 ns, respectively. The light output spectrum recorded by the SCI-5550W software is shown in Fig.~\ref{fig:cs137_raw} and the corresponding calibrated spectra is shown in Fig.~\ref{fig:cs137}. The Compton edge is clearly visible on the Cs-137 light output spectrum.


\subsection{PSD}
\begin{figure}[!htbp]
    \centering
    \includegraphics[width=\linewidth]{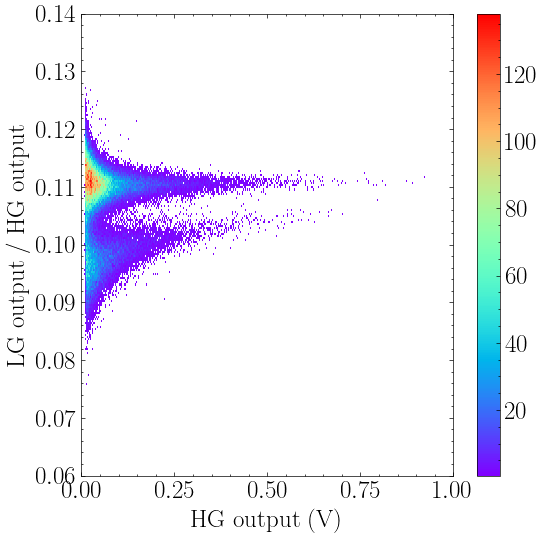}
    \caption{ASIC-based PSD scatter density plot. Pixels are colored by number of counts.}
    \label{fig:asic-psd}
\end{figure}
\begin{figure}[!htbp]
    \centering
    \includegraphics[width=\linewidth]{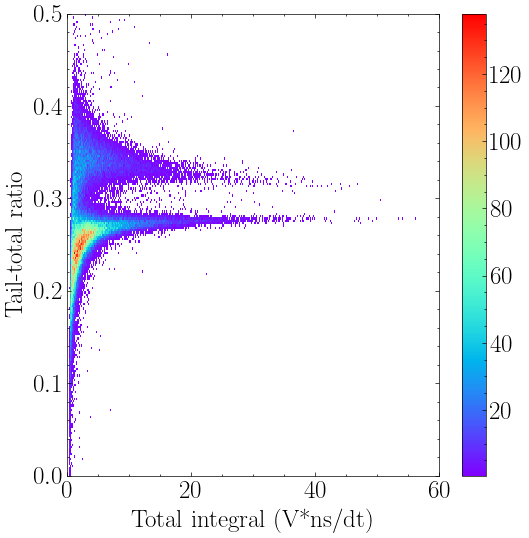}
    \caption{Charge-integration based PSD scatter density plot. Pixels are colored by number of counts.}
    \label{fig:ci-psd}
\end{figure} 

\added{We sent the SiPM signals in the regular stilbene dataset introduced in Section~\ref{section:PSD} to the simulated FDTF and calculated the LG and HG outputs for each signal. The PSD parameter, defined as the ratio between LG and HG outputs, was calculated for each pulse} and a scatter density plot of the PSD parameter is shown in Fig.~\ref{fig:asic-psd}. For comparison, we also show in Fig.~\ref{fig:ci-psd} the PSD scatter density plot obtained by using the traditional charge-integration method, with the short gate and long gate set to 500~ns and 2000~ns, respectively. In Fig.~\ref{fig:asic-psd}, the two pulse types can be clearly discriminated, with the upper band being gamma ray pulses, and the lower band being neutron pulses. The separation between the two types of pulses is comparable. Therefore, it is possible to perform PSD utilizing existing readout electronics on the A55CIT4 board.

\section{Conclusions and Future work}
In this work, we have designed a imaging system using MicroFJ-30020 SiPM arrays and a A55CIT4+DT5550W readout system. The system demonstrates good gamma ray responses. We also showed in simulation that it's possible to perform ASIC-based PSD using available readout electronics on the A5CIT4 board. Next, we will use the system to measure a PuBe neutron source and experimentally test the PSD capability.




\bibliographystyle{IEEEtran}
\bibliography{references}
%






\end{document}